# Forecasting NIFTY 50 benchmark Index using Seasonal ARIMA time series models

Amit Tewari

H. Milton Stewart School of Industrial and Systems Engineering (ISyE)

Georgia Institute of Technology

North Ave NW, Atlanta, GA 30332, United States

(404) 894-2000

E-mail: atewari35@gatech.edu

## Abstract

This paper analyses how Time Series Analysis techniques can be applied to capture movement of an exchange traded index in a stock market. Specifically, Seasonal Auto Regressive Integrated Moving Average (SARIMA) class of models is applied to capture the movement of Nifty 50 index which is one of the most actively exchange traded contracts globally [1]. A total of 729 model parameter combinations were evaluated and the most appropriate selected for making the final forecast based on AIC criteria [8]. NIFTY 50 can be used for a variety of purposes such as benchmarking fund portfolios, launching of index funds, exchange traded funds (ETFs) and structured products. The index tracks the behaviour of a portfolio of blue chip companies, the largest and most liquid Indian securities and can be regarded as a true reflection of the Indian stock market [2].

## Keywords

Time Series Analysis, Stock Market Index, Nifty 50, Seasonal ARIMA, SARIMA, NSE

## Introduction

**NIFTY 50**

The NIFTY 50 index is National Stock Exchange of India's benchmark broad based stock market index for the Indian equity market. Full form of NIFTY is National Stock Exchange Fifty. It represents the weighted average of 50 Indian company stocks in 13 sectors and is one of the two main stock indices used in India, the other being the BSE Sensex.

NIFTY 50 is also one of the world's most actively traded contract [1].

In the United States, the term Nifty Fifty was an informal designation for fifty popular large-cap stocks on the New York Stock Exchange in the 1960s and 1970s that were widely regarded as solid buy and hold growth stocks, or "Blue-chip" stocks. These fifty stocks are credited by historians with propelling the bull market of the early 1970s [3, 4].

**ARIMA model**

Auto Regressive Integrated Moving Average (ARIMA) [5] is a class of models that 'explains' a given time series based on its own past values, that is, its own lags and the lagged forecast errors, so that equation can be used to forecast future values.

Any 'non-seasonal' time series that exhibits patterns and is not a random white noise can be modelled with ARIMA models.

An ARIMA model is characterized by 3 terms: p, d, q where,

- p is the order of the AR term
- d is the number of differencing required to make the time series stationary
- q is the order of the MA term

A pure Auto Regressive (AR only) model is one where $Y_t$ depends only on its own lags. That is, $Y_t$ is a function of the 'lags of $Y_t$'.

$$Y_t = \alpha + \beta_1 Y_{t-1} + \beta_2 Y_{t-2} + \cdots \beta_p Y_{t-p} + \epsilon_1$$

A pure Moving Average (MA only) model is one where $Y_t$ depends only on the lagged forecast errors.

$$Y_t = \alpha + \epsilon_t + \phi_1 \epsilon_{t-1} + \phi_2 \epsilon_{t-2} + \cdots \phi_q \epsilon_{t-q}$$

Combining above we get,

$$Y_t = \alpha + \beta_1 Y_{t-1} + \beta_2 Y_{t-2} + \cdots \beta_p Y_{t-p} + \epsilon_t + \phi_1 \epsilon_{t-1} + \phi_2 \epsilon_{t-2} + \cdots \phi_q \epsilon_{t-q}$$

Predicted $Y_t$ = Constant + Linear combination Lags of Y (upto p lags) + Linear Combination of Lagged forecast errors (upto q lags)

**Seasonal ARIMA (SARIMA)**

Many previous studies have focussed on using ARIMA models for forecasting stock markets [9]. However, a shortcoming of plain ARIMA model is that it does not support seasonality. If the time series has defined seasonality, then, SARIMA which uses seasonal differencing is a more appropriate model in which values from previous season are subtracted instead of subtracting consecutive terms [6].

SARIMA can be represented by SARIMA(p,d,q) x (P,D,Q,m), where,

- (p,d,q) are non-seasonal terms of the model

- (P,D,Q,m) are seasonal autoregressive (SAR), order of seasonal differencing (D) and seasonal moving average (SMA) and frequency of the time series terms respectively.

## Methodology and Results

Last 11 years data (2009-2019) for this research was downloaded from NSE website [2]. Data post-2008 is only used in this research since Indian economy was coming out of a major financial recession during 2008 and was going through systematic and structural changes.

Fig.1 represents Nifty 50 index closing price (monthly average) time series data from Jan 2011 to Dec 2019. Fig. 2 shows boxplots of Nifty 50 closing prices year on year. In Fig. 3, I plotted the time series but this time months on x-axis. Next the time series was decomposed into trend, seasonality and noise components (Fig. 4).

As first step, time series model is created using data from 2009-2017. Period Jan 2018 – Dec 2018 is used for one step ahead forecasting (Fig. 5). Period Jan 2019 – June 2019 is reserved for model validation by comparing model forecasted values with actual observations (Fig. 6). For the 6 month hold-old validation period MAPE was 0.9%. RMSE was 139.67 for actual values in range [10800, 11800]. Other common error metrics are provided in Table 1. Next, time series model is created using entire data range (2009-2019) to generate future forecasts for Jan 2020 – Dec 2020 (Fig. 7). Model parameters selection in all cases was done using AIC criteria [8]. The model with minimum AIC score (AIC=1008.06) was selected as the final model for future forecasts (Fig. 8).

## Discussion

We observed an upwards trend in Fig. 1 which is reinforced in Fig. 2 which shows boxplots of Nifty 50 closing prices year on year. In Fig. 3 we also get a hint of seasonality. From Jan to Oct there is a slight increase and thereafter a slight decrease from Oct to Dec for most of the years. In the decomposed time series (Fig. 4) we can clearly observe seasonality (12 months) as well as cyclicity (approximately 4 year cycles). Thus we can see that the time series can be decomposed into trend, seasonality and noise so we can use Seasonal ARIMA model for this time series data to generate future forecasts.

For the 6 month hold-old validation period MAPE of 0.9% implies the created model was about 99.1% accurate in predicting the next 6 month forecasts. Likewise RMSE of 139.67 in actual range [10800, 11800] also suggest a useful model. Once it is established that SARIMA(2,2,1)x(2,2,1,12) can be a useful model for this time series data, it is used to forecast future values for Jan 2020 – Dec 2020 (Fig. 7) using entire available data (2009-2019) for modelling.

## Conclusion

In this paper, I have tried to forecast the stock price movement of the Nifty 50 index which is the largest and most liquid Indian securities. Seasonal ARIMA (SARIMA) model parameters were discovered using AIC criteria. To apply the time series techniques, I used 11 year data (2009-2019) for this research downloaded from NSE website. Data post-2008 was only used in this research since Indian economy was coming out of a major financial recession during 2008 and was going through systematic and structural changes.

The stationarity in this time series has been removed using decomposing and differencing. After that, Seasonal ARIMA (SARIMA) parameters - p, d, q, P, D, Q, m were determined based on AIC criteria and using that model average index price movement for next 12 months forecasted and plotted. There can also be some uncertainties associated in forecasting stock and index price movements e.g. due to macro as well as micro-economic factors which can influence price movement. However these are outside the scope of the current research and methodology.

# References

[1] http://www.world-exchanges.org/our-work/articles/wfe-ioma-2018-derivatives-report

[2] https://www.nseindia.com/index_nse.htm

[3] https://en.wikipedia.org/wiki/NIFTY_50

[4] https://en.wikipedia.org/wiki/Nifty_Fifty

[5] G.E.P. Box, G. Jenkins, "Time Series Analysis, Forecasting and Control", Holden-Day, San Francisco, CA, 1970

[6] https://otexts.com/fpp2/seasonal-arima.html

[7] https://www.statsmodels.org/dev/generated/statsmodels.tsa.statespace.sarimax.SARIMAX.html

[8] Akaike, H. (1973), "Information theory and an extension of the maximum likelihood principle", in Petrov, B. N.; Csáki, F. (eds.), 2nd International Symposium on Information Theory, Tsahkadsor, Armenia, USSR, September 2-8, 1971, Budapest: Akadémiai Kiadó, pp. 267–281.

[9] Ariyo, A. A., Adewumi, A. O., and Ayo, C. K. (2014). Stock price prediction using the arima model. In Computer Modelling and Simulation (UKSim), 2014 UKSim-AMSS 16th International Conference on, pages 106–112. IEEE.

# Figures

**Figure 1**

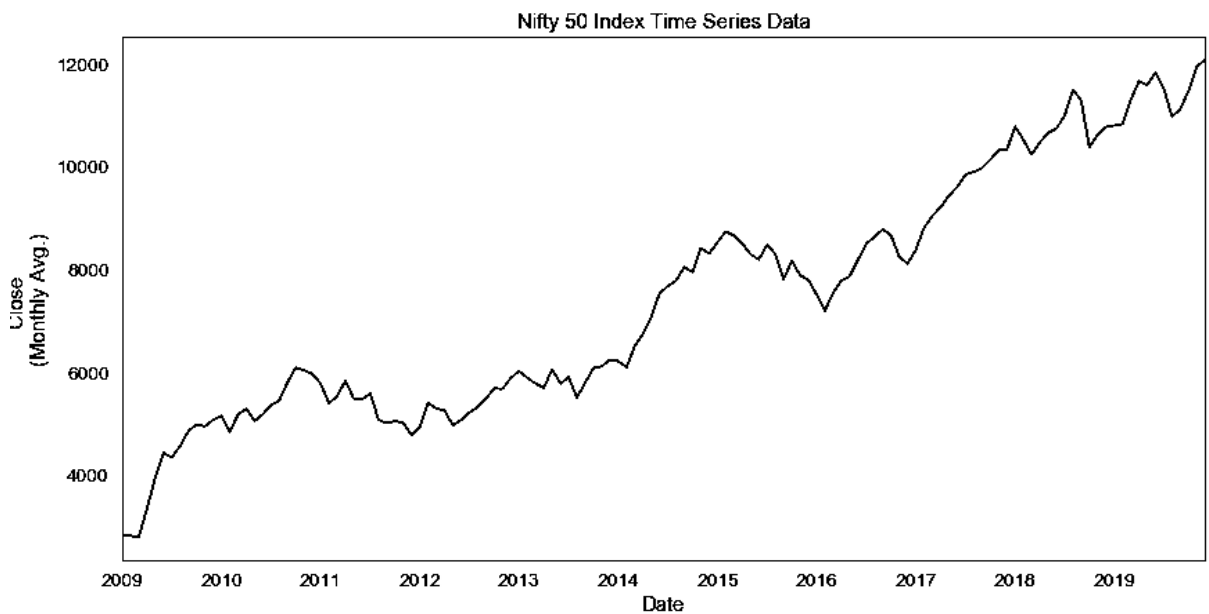

**Figure 2**

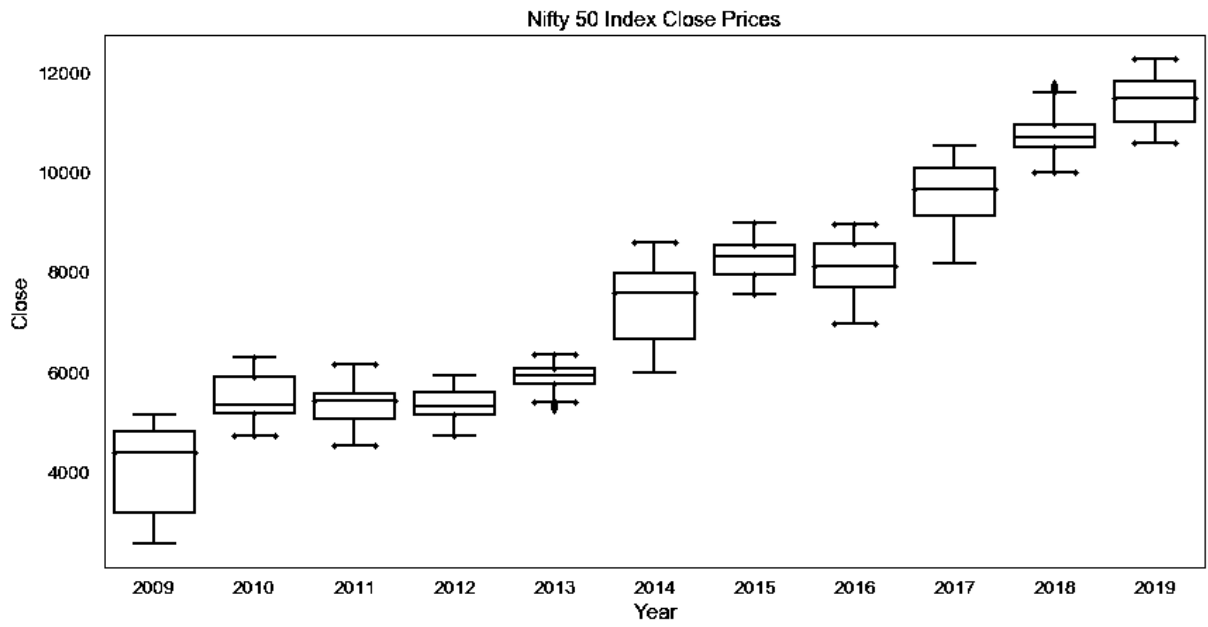

**Figure 3**

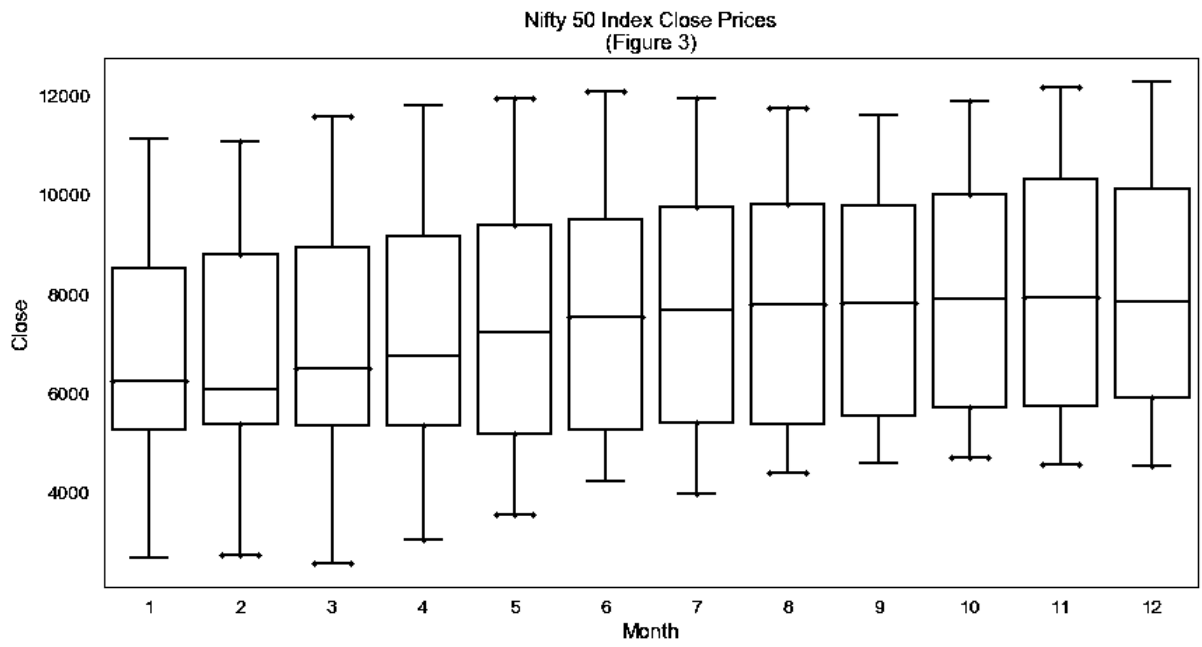

**Figure 4**

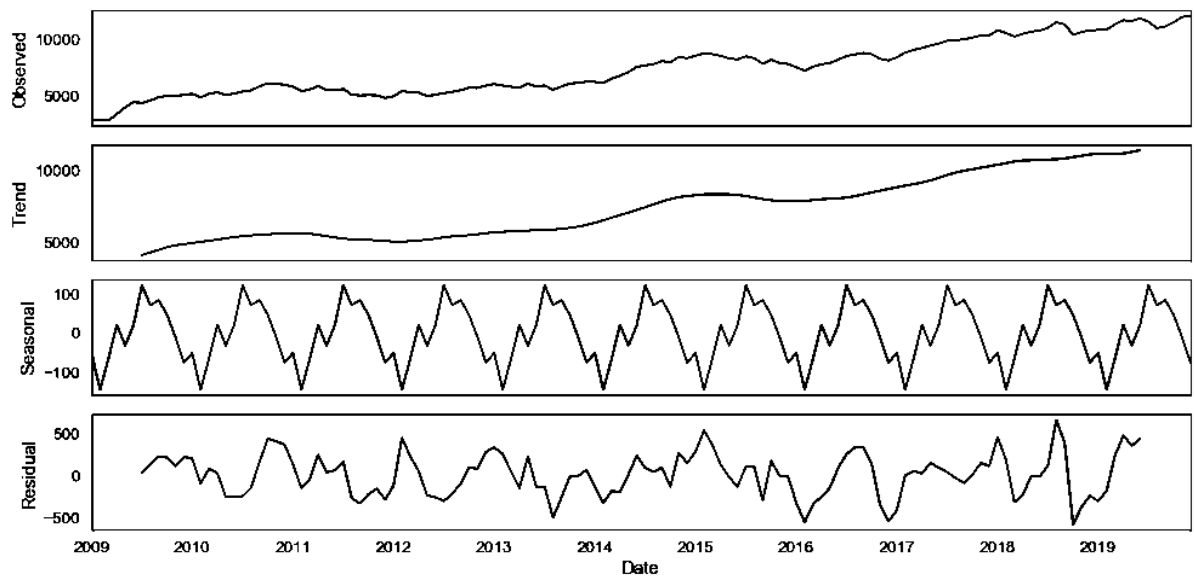

**Figure 5**

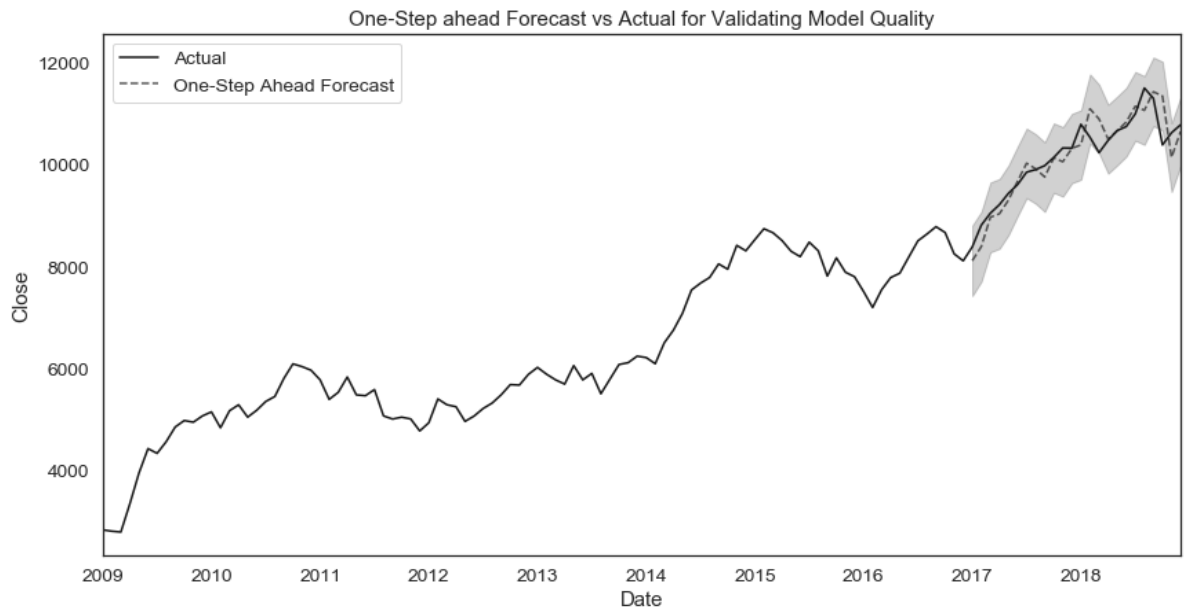

**Figure 6**

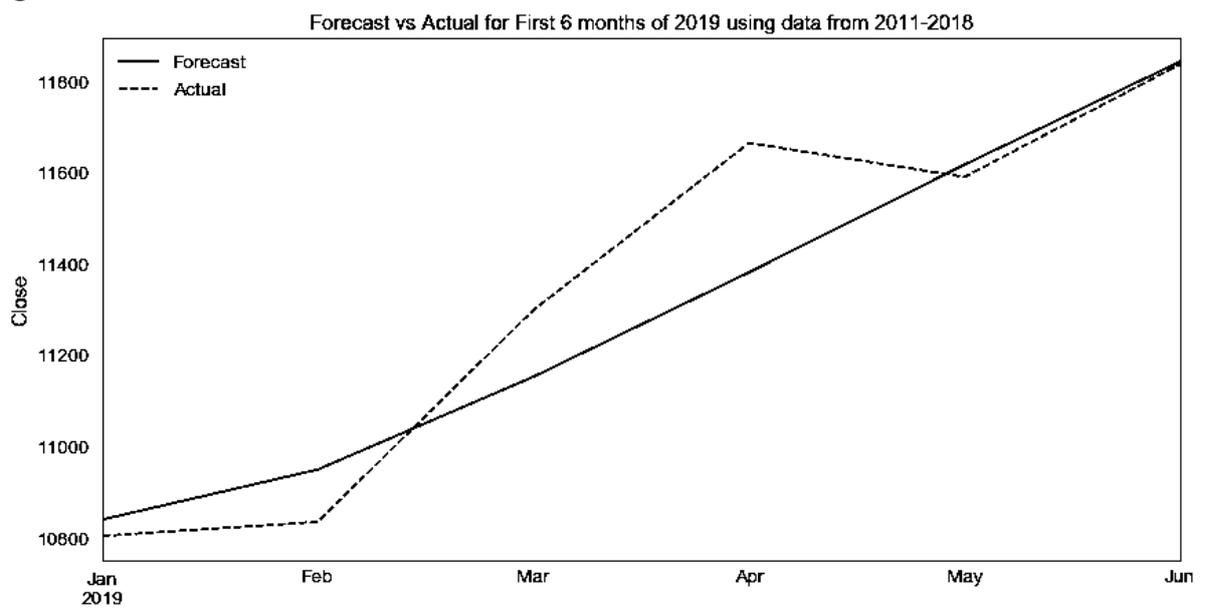

**Figure 7**

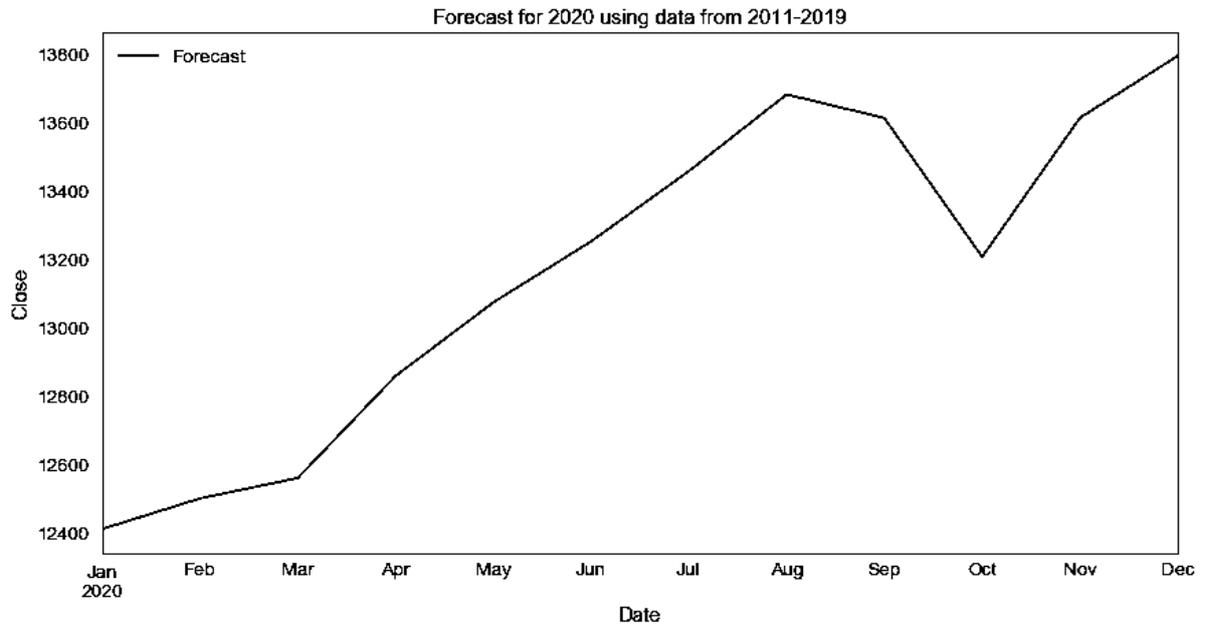

**Figure 8**

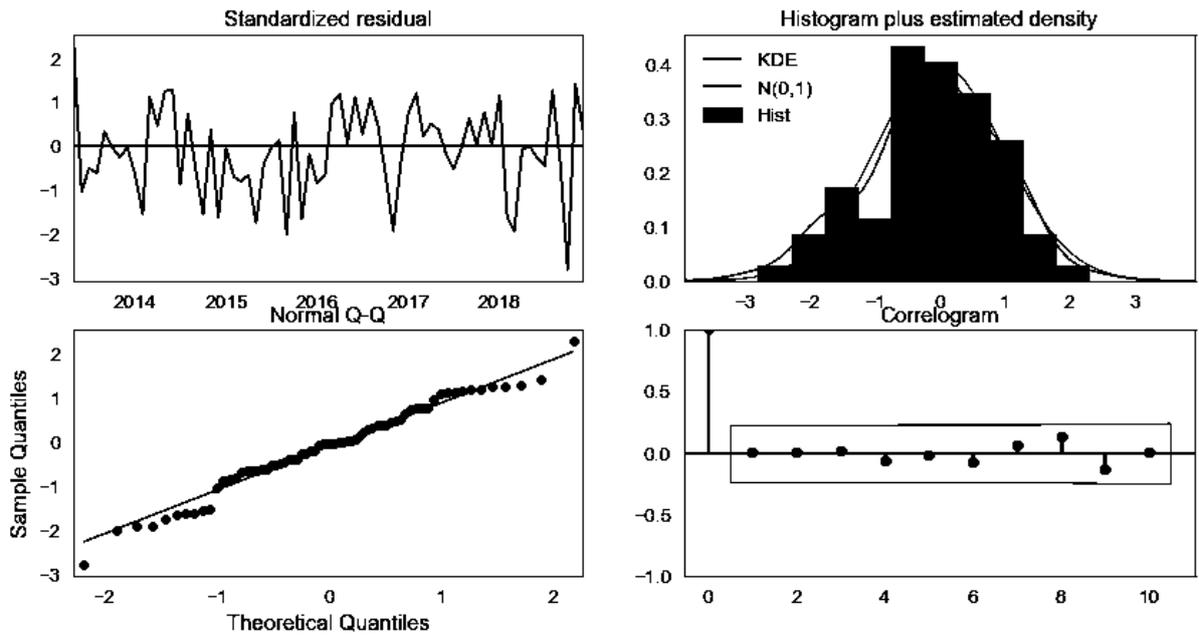

**Tables**

**Table 1**

```
==============================================================================
                 coef    std err          z      P>|z|      [0.025      0.975]
------------------------------------------------------------------------------
ar.L1         -0.1151      0.147     -0.784      0.433      -0.403       0.173
ar.L2         -0.2040      0.160     -1.274      0.203      -0.518       0.110
ma.L1         -0.8222      0.101     -8.114      0.000      -1.021      -0.624
ar.S.L12      -0.8013      0.140     -5.728      0.000      -1.076      -0.527
ar.S.L24      -0.3174      0.133     -2.388      0.017      -0.578      -0.057
ma.S.L12      -1.0004      0.188     -5.311      0.000      -1.370      -0.631
sigma2      1.025e+05   1.84e-06   5.59e+10      0.000    1.03e+05    1.03e+05
==============================================================================
```

**Table 2**

| | | |
|---|---|---|
| Mean Absolute Percentage Error | MAPE | 0.89 |
| Mean Error | ME | -41.06 |
| Mean Absolute Error | MAE | 102.23 |
| Mean Percentage Error | MPE | -0.34 |
| Root Mean Square Error | RMSE | 139.67 |